\documentclass[11pt]{article}
\usepackage{typearea}\typearea{13}
\usepackage{epsfig,amsmath,amsfonts,amssymb,upgreek,dsfont}
\newtheorem{theorem}{Theorem}[section]
\newtheorem{lemma}[theorem]{Lemma}

\makeatletter
\@addtoreset{equation}{section}
\makeatother

\makeatletter
\long\def\@makecaption#1#2{{\small
\advance\leftskip1cm
\advance\rightskip1cm
\vskip\abovecaptionskip
\sbox\@tempboxa{#1: #2}%
\ifdim \wd\@tempboxa >\hsize
 #1: #2\par
\else
\global \@minipagefalse
\hb@xt@\hsize{\hfil\box\@tempboxa\hfil}%
\fi
\vskip\belowcaptionskip}}
\makeatother
\def\eq#1\en{\begin{equation}#1\end{equation}}  
\def\eqa#1\ena{\begin{align}#1\end{align}}
\def\eqg#1\eng{\begin{gather}#1\end{gather}}
\newcommand{\lb}[1]{\label{e:#1}}
\newcommand{\rlb}[1]{\eqref{e:#1}} 
\newcommand{\nl}{\notag\\}
\newcommand{\itext}{\notag\intertext}

\newcommand{\qedm}{\rule{1.5mm}{3mm}}

\newcommand{\snorm}[1]{\Vert#1\Vert}

\newcommand{\sbkt}[1]{\langle#1\rangle}

\newcommand{\bra}[1]{\langle#1|}
\newcommand{\ket}[1]{|#1\rangle}

\newcommand{\sumtwo}[2]%
{\mathop{\sum_{#1}}_{#2}}
\newcommand{\sumthree}[3]%
{\mathop{\mathop{\sum_{#1}}_{#2}}_{#3}}
\newcommand{\sumfour}[4]%
{\mathop{\mathop{\mathop{\sum_{#1}}_{#2}}_{#3}}_{#4}} 
\newcommand{\prodtwo}[2]%
{\mathop{\prod_{#1}}_{#2}}
\newcommand{\mintwo}[2]%
{\mathop{\min_{#1}}_{#2}}
\newcommand{\maxtwo}[2]%
{\mathop{\max_{#1}}_{#2}}
\newcommand{\maxthree}[3]%
{\mathop{\mathop{\max_{#1}}_{#2}}_{#3}}
\newcommand{\limtwo}[2]%
{\mathop{\lim_{#1}}_{#2}}
\newcommand{\suptwo}[2]%
{\mathop{\sup_{#1}}_{#2}}
\newcommand{\supthree}[3]%
{\mathop{\mathop{\sup_{#1}}_{#2}}_{#3}}
\newcommand{\supfour}[4]%
{\mathop{\mathop{\mathop{\sup_{#1}}_{#2}}_{#3}}_{#4}} 
\newcommand{\inftwo}[2]%
{\mathop{\inf_{#1}}_{#2}}
\newcommand{\infthree}[3]%
{\mathop{\mathop{\inf_{#1}}_{#2}}_{#3}}
\newcommand{\inffour}[4]%
{\mathop{\mathop{\mathop{\inf_{#1}}_{#2}}_{#3}}_{#4}} 
\newcommand\calA{{\cal A}}

\newcommand\calH{{\cal H}}

\newcommand\calN{{\cal N}}

\newcommand\calW{{\cal W}}

\newcommand{\ha}{\hat{a}}
\newcommand{\had}{\hat{a}^\dagger}
\newcommand{\hb}{\hat{b}}
\newcommand{\hbd}{\hat{b}^\dagger}
\newcommand{\hc}{\hat{c}}
\newcommand{\hcd}{\hat{c}^\dagger}

\newcommand{\hu}{\hat{u}}

\newcommand{\hn}{\hat{n}}

\newcommand{\hA}{\hat{A}}
\newcommand{\hB}{\hat{B}}

\newcommand{\hH}{\hat{H}}

\newcommand{\hN}{\hat{N}}
\newcommand{\hP}{\hat{P}}

\newcommand{\bbQ}{\mathbb{Q}}
\newcommand{\bbR}{\mathbb{R}}
\newcommand{\bbZ}{\mathbb{Z}}
\newcommand{\ep}{\epsilon}

\newcommand{\up}{\uparrow}
\newcommand{\dn}{\downarrow}

\newcommand{\Di}{\mathit{\Delta}}

\newcommand{\Tr}{\operatorname{Tr}}

\usepackage{color}
\usepackage[normalem]{ulem}
\definecolor{fluorescentpink}{rgb}{1.0, 0.08, 0.58}
\definecolor{forestgreen}{rgb}{0.13, 0.55, 0.13}

\newcommand{\La}{\Lambda}
\newcommand{\kz}{\ket{0}}
\newcommand{\ko}{\ket{1}}
\newcommand{\epo}{\epsilon_0}
\newcommand{\Du}{\Di u}
\newcommand{\Duo}{\Du_0}
\newcommand{\DN}{\Di N}
\newcommand{\Pneq}{\hP_\mathrm{neq}}

\newcommand{\kv}{\ket{\Phi_\mathrm{vac}}}
\newcommand{\bv}{\bra{\Phi_\mathrm{vac}}}
\newcommand{\kPhi}{\ket{\Phi}}

\newcommand{\kPsi}{\ket{\Psi}}
\newcommand{\bPsi}{\bra{\Psi}}
\newcommand{\kPz}{\ket{\Phi(0)}}
\newcommand{\kPt}{\ket{\Phi(t)}}
\newcommand{\bPz}{\bra{\Phi(0)}}
\newcommand{\bPt}{\bra{\Phi(t)}}
\newcommand{\Hmc}{\calH_\mathrm{mc}}
\newcommand{\Lh}{L+\frac{1}{2}}
\newcommand{\summu}{\sum_{\mu=1}^{p-1}}

\newcommand{\la}{\lambda}
\newcommand{\hd}{\hat{d}}
\newcommand{\hdd}{\hat{d}^\dagger}

\newcommand{\uo}{u^\mathrm{out}}
\newcommand{\Proj}{\operatorname{Proj}}

\begin{document}
\renewcommand{\thefootnote}{\fnsymbol{footnote}}

\begin{flushright}
\footnotesize
technical note not (yet) submitted to any journals, April 2024.
\end{flushright}
\noindent
{\Large\bf Heat flows from hot to cold}

\vspace{1mm}
\noindent
{\large\bf A simple rigorous example of thermalization in an isolated macroscopic quantum system}

\medskip\noindent
Hal Tasaki\footnote{%
Department of Physics, Gakushuin University, Mejiro, Toshima-ku, 
Tokyo 171-8588, Japan.
}
\renewcommand{\thefootnote}{\arabic{footnote}}
\setcounter{footnote}{0}

\begin{quotation}
\small\noindent
In the present note, we discuss a simple example of a macroscopic quantum many-body system in which the approach to thermal equilibrium from an arbitrary initial state in the microcanonical energy shell is proved without relying on any unproven assumptions.
The model, which is equivalent to a free fermion chain, is designed to be a toy model for a weakly heat-conducting one-dimensional solid.
We take a phenomenological point of view and perceive that the system is in thermal equilibrium when the measured coarse-grained energy distribution is uniform.

The result on thermalization reported here is a variation (and an improvement) of our previous result on the irreversible expansion in a free fermion chain \cite{TasakiFreeFermion}.
As far as we know, this is the first concrete and rigorous realization of the philosophy on the foundation of equilibrium statistical mechanics proposed by von Neumann in 1929 \cite{vonNeumann,GLTZ}, and further developed recently by Goldstein, Lebowitz, Mastrodonato, Tumulka, and Zangh\`\i~\cite{GLMTZ09b} and the present author \cite{Tasaki2010,Tasaki2016}, namely, to characterize thermal equilibrium from a macroscopic viewpoint and to make use of the strong ETH \cite{Deutsch1991,Srednicki1994,Tasaki1998,RigolSrednicki2012,DAlessioKafriPolkovnikovRigol2016} to control the long-time dynamics.

\medskip\noindent
{\em This note will be the most technical part of my longer article on thermalization, ``What is thermal equilibrium and how do we get there?''.
I am making this document public at this stage since I have already announced (and will announce) the results at some of my talks.
}

\end{quotation}


\tableofcontents

\section{Problem and the thermodynamic point of view}
\label{s:intro}

In this note, we focus on the phenomenon of thermalization in a one-dimensional uniform solid that weakly conducts heat.

\begin{figure}
\centerline{\epsfig{file=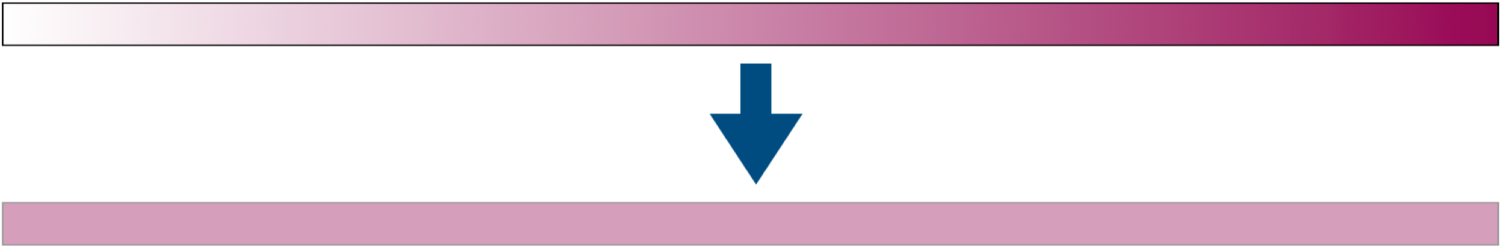,width=10truecm}}
\caption[dummy]{
We study thermalization in a weakly heat-conducting one-dimensional solid isolated from the external environment.
It is one of the fundamental premises of thermodynamics that the solid reaches a unique thermal equilibrium state after a sufficiently long time.
No matter how the energy is distributed in the initial state, the thermal equilibrium state is characterized by a uniform energy distribution.
We shall study this problem in a simple microscopic quantum model and prove, without relying on any unjustified assumptions, that such thermalization takes place.
}
\label{f:heat}
\end{figure}

Let us first examine this problem from the macroscopic thermodynamic point of view.
Suppose the solid is initially in an arbitrary state with total (macroscopic) energy $U_0$ and is completely isolated from the external environment.
Then, it is a fundamental premise of thermodynamics that, after a sufficiently long time, the solid will approach a unique thermal equilibrium state in which the energy is equally distributed over the solid with the uniform energy density $u_0$.
See Figure~\ref{f:heat}.

\begin{figure}
\centerline{\epsfig{file=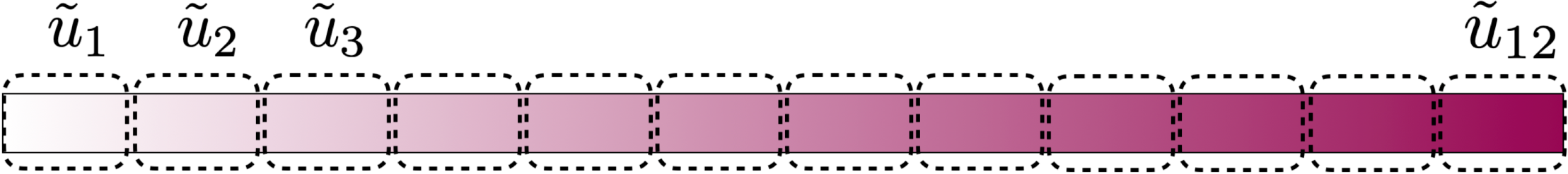,width=10truecm}}
\caption[dummy]{
We divide our solid into $m$ identical pieces and focus on the energy densities in these pieces.
Here we set $m=12$.
Initially, the $m$ pieces may have arbitrary energy densities, which we here wrote $\tilde{u}_1,\ldots,\tilde{u}_{12}$.}
\label{f:dec1}
\end{figure}

\begin{figure}
\centerline{\epsfig{file=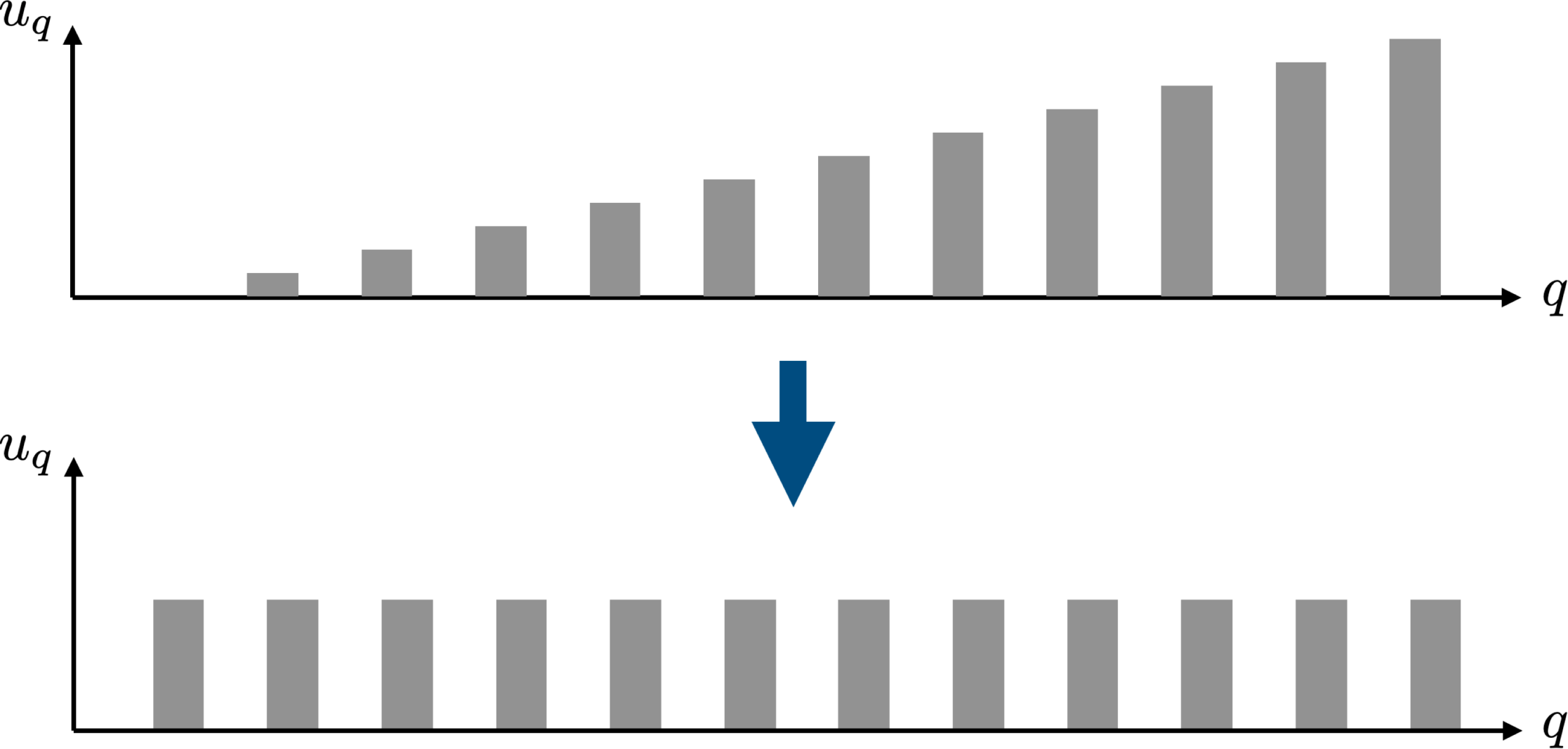,width=12truecm}}
\caption[dummy]{
The energy densities in the pieces $q=1,\ldots,m$.
In the initial nonequilibrium state of Figure~\ref{f:dec1}, the energy densities take nonuniform values.
In the thermal equilibrium state, all the pieces have the same energy density $u_0$, which is determined by the total energy in the initial state.
This phenomenological and operational characterization of thermal equilibrium will be used throughout the present note.
}
\label{f:histogramas}
\end{figure}

To make this idea more quantitative, we (fictitiously) divide the solid into $m$ pieces with identical lengths and focus on the energy densities in these pieces.
See Figure~\ref{f:dec1}.
Here, $m$ may note be small, but it is assumed that each piece is still a macroscopic thermodynamic system.
In the initial nonequilibrium state, the energy densities in the $m$ pieces may take arbitrary values, $\tilde{u}_1,\ldots,\tilde{u}_m$, but they all take the identical value $u_0$ in the equilibrium state.
See Figure~\ref{f:histogramas}.

Throughout the present note, we shall take a phenomenological and operational point of view when characterizing states of the solid.
This means we focus solely on what a macroscopic observer who is ignorant of microscopic physics finds out.
To be specific, we assume that our macroscopic observer pays attention only to the coarse-grained energy distribution in the solid as formulated above.
In particular, if the observer finds the energy densities in all the pieces to be $u_0$, then she/he concludes that the solid is in thermal equilibrium.
We shall use this definition of thermal equilibrium even when we study microscopic physics in the following sections.\footnote{%
\label{fn:contextuality}%
Note that our definition of thermal equilibrium thus depends on the choice of number $m$.
One may say that our notion of thermal equilibrium is not only phenomenological (in the sense we focus only on the results of macroscopic observations) but also contextual (in the sense the notion may depend on which quantities the observer measures).
It should be emphasized that contextuality never implies that the notion is subjective.
}

The goal of the present note is to establish the presence of thermalization in the above phenomenological sense in a simple microscopic model for our one-dimensional solid.

\section{Quantum mechanical model and main results on thermalization}
\label{s:main}
Here, we define our simple model of solid (section~\ref{s:def}), discuss our phenomenological and operational criterion for thermal equilibrium (section~\ref{s:criterion}), and state the main theorems for thermalization (section~\ref{s:maintheorems}).
We then introduce two essential Lemmas and see how the theorems can be proved assuming these lemmas (section~\ref{s:proof1}).
Finally we briefly discuss the puzzling issue about integrability and thermalization (section~\ref{s:integrable}).

\subsection{Definition}
\label{s:def}
\begin{figure}
\centerline{\epsfig{file=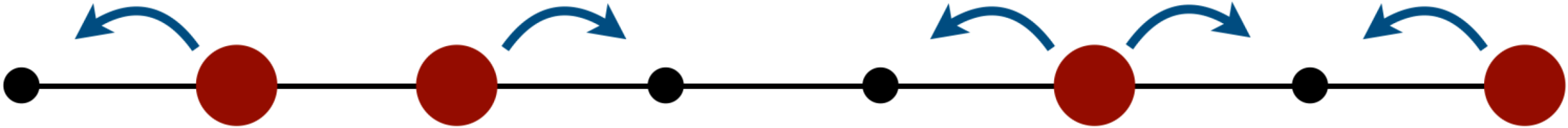,width=12truecm}}
\caption[dummy]{
The simple model of a one-dimensional solid with $L=8$.
Each atom can be in the zero energy state (denoted by the small dot) or in the excited state (denoted by the big dot) with energy $\epo>0$.
An excitation can hop to a neighboring site with a small amplitude $\eta\epo/2$.
}
\label{f:model1}
\end{figure}

Let us consider a ``solid'' consisting of $L$ atoms indexed by $x\in\{1,\ldots,L\}$ and forming a chain.
We assume that each atom can take two states, namely, a state with zero energy and an excited state with energy $\epo>0$.
We describe the state of the $x$-th atom by the variable $\mu_x=0,1$, where 0 and 1 represent the zero-energy state and the excited state, respectively.
The sum of the energies of all atoms is then given by $\epo\sum_{x=1}^L\mu_x$.
See Figure~\ref{f:model1}.

In the quantum mechanical notation, we denote the zero-energy and the excited states of the $x$-th atom as $\kz_x$ and $\ko_x$, respectively.
We assume that the set $\{\kz_x,\ko_x\}$ forms an orthonormal basis of the two-dimensional Hilbert space of the $x$-th atom.
The $2^L$ dimensional Hilbert space $\calH$ of the whole system is the tensor product of the Hilbert spaces of all atoms.
It is spanned by the basis states
\eq
\ket{\mu_1,\ldots,\mu_L}=\bigotimes_{x=1}^L\ket{\mu_x}_x,
\lb{basis}
\en
with $\mu_x=0,1$.

We define the lowering and the raising operators $\ha_x$ and $\had_x$ acting on the local Hilbert space of the $x$-th atom by 
\eq
\ha_x\ko_x=\kz_x,\quad \ha_x\kz_x=0,\quad \had_x\ko_x=0,\quad  \had_x\kz_x=\ko_x.
\en
The number operator defined as $\hn_x=\had_x\ha_x$ satisfies $\hn_x\ko_x=\ko_x$ and $\hn_x\kz_x=0$.
Note that these operators for different atoms, i.e., different indices $x$, commute with each other.

The Hamiltonian of our solid is
\eq
\hH=\epo\sum_{x=1}^L\hn_x+\frac{\eta\epo}{2}\Bigl\{\sum_{x=1}^{L-1}(\had_x\ha_{x+1}+\had_{x+1}\ha_x)-\hn_L\Bigr\}.
\lb{HS}
\en
The first term on the right-hand side is the sum of the energies of atoms.
The second term represents processes in which an excitation is transferred to a neighboring site with amplitude $\eta\epo/2$.
We added an on-site potential at site $L$ in order to break the reflection symmetry and guarantee that the model is free from degeneracy.  See Lemma~\ref{L:nondeg} below.
We assume that $\eta$ is small, i.e., the transfer energy is much smaller than the excitation energy.

Let us introduce the microcanonical energy shell, which plays an important role in the present approach.
It is a subspace of $\calH$ defined by
\eq
\Hmc=\operatorname{span}\Bigl\{\kPsi\,\Bigl|\,\hH\kPsi=E\kPsi, \Bigl|\frac{E}{L}-u_0\Bigr|\le \Duo\Bigr\},
\lb{Hmc}
\en
where $u_0\in(0,\epo)$ is a fixed energy density and $\Duo$ defines the small energy width that is not perceived by a macroscopic observer.
We thus assume $\Duo/u_0$ is small.
At the same time, $\Duo$ must be sufficiently large so that the dimension of $\Hmc$ is large.

We interpret the energy shell $\Hmc$ as a collection of states with a definite total macroscopic energy $U_0=u_0L$ in the thermodynamic setup.
We also note that the microcanonical average is nothing but the normalized trace in $\Hmc$, i.e.,
\eq
\sbkt{\cdots}_\mathrm{mc}=\frac{\Tr_{\Hmc}[\cdots]}{\Tr_{\Hmc}[1]}.
\en

\subsection{Criterion for thermal equilibrium}
\label{s:criterion}
We shall formulate our phenomenological and operational notion of thermal equilibrium discussed in section~\ref{s:intro} in the language of microscopic quantum mechanics.
Our notion may be viewed as a variant of MATE (macroscopic thermal equilibrium) formulated in \cite{GoldsteinHuseLebowitzTumulka2015}.

Let $m\in\{2,3,\ldots\}$, and assume that $L$ is divisible by $m$.\footnote{%
We make this assumption only for notational simplicity.
}
We decompose the chain into $m$ identical parts as
\eq
\{1,\ldots,L\}=\bigcup_{q=1}^m\La_q,
\en
where
\eq
\La_q=\Bigl\{\frac{L}{m}(q-1)+1,\ldots,\frac{L}{m}q\Bigr\},
\lb{Laq}
\en
is the sublattice with $L/m$ sites.
We then define for $q=1,\ldots,m$ the Hamiltonian on $\La_q$ by 
\eq
\hH_q=\epo\sum_{x\in\La_q}\hn_x+\frac{\eta\epo}{2}\sum_{x=\frac{L}{m}(q-1)+1}^{\frac{L}{m}q-1}(\had_x\ha_{x+1}+\had_{x+1}\ha_x).
\lb{Hq}
\en
Note that different $\hH_q$ commute because they do not contain operators on common sites.
Consequently, the sum $\sum_{q=1}^m\hH_q$ is slightly different from the total Hamiltonian $\hH$.

As we discussed in section~\ref{s:intro}, we assume that the macroscopic observer pays attention only to the energy densities in $\La_q$.
It is then convenient to define the energy density operator
\eq
\hu_q=\frac{m}{L}\hH_q,
\lb{uq}
\en
for $q=1,\ldots,m$.
It, of course, holds that
\eq
\sbkt{\hu_q}_\mathrm{mc}\simeq u_0.
\en

In the quantum mechanical setting, our observer is given an arbitrary pure state $\kPhi\in\Hmc$.
It corresponds to a state with a definite macroscopic energy $U_0=u_0L$, or the average energy density $u_0$.
The observer makes a simultaneous projective measurement of the mutually commuting operators $\hu_1,\ldots,\hu_m$ and obtains a set of outcomes $\uo_1,\ldots,\uo_m$.
Then the observer concludes that the solid is in thermal equilibrium if the measurement result satisfies $|\uo_q-u_0|\le\Du$ for all $q=1,\ldots,m$, where $\Du>0$ is a specified (small) precision.
The precision $\Du$ should be macroscopically observable, and hence we assume $\Du$ is much larger than $\Duo$.\footnote{%
This criterion of thermal equilibrium depends on the choice of the division number $m$ and the precision $\Du$.
As we discussed in footnote~\ref{fn:contextuality}, our notion of thermal equilibrium is contextual.
}

This scenario motivates us to define the projection operator onto nonequilibrium space, the nonequilibrium projection for short, as\footnote{%
Here is a more careful construction.
Since $\hu_1,\ldots,\hu_m$ commute, one can form an orthonormal basis of the whole Hilbert space $\calH$ that consists of their simultaneous eigenstates $\ket{\Xi_\nu}$ with $\nu=1,\ldots,2^L$.
Suppose that $\hu_q\ket{\Xi_\nu}=u^{(\nu)}_q\ket{\Xi_\nu}$ for $q=1,\ldots,m$, and let $\calN\subset\{1,\ldots,2^L\}$ be the set of $\nu$ such that $|u^{(\nu)}_q-u_0|\ge\Du$ for some $q$.
Then we define $\Pneq=\sum_{\nu\in\calN}\ket{\Xi_\nu}\bra{\Xi_\nu}$.
}
\eq
\Pneq=\Proj\bigl[\,|\hu_q-u_0|\ge\Du\ \text{for some $q\in\{1,\ldots,m\}$}\,\bigr].
\lb{Pneq}
\en
The expectation value of $\Pneq$ gives the probability that the state in question is {\em not}\/ found in thermal equilibrium.
Therefore, if it holds for a normalized state $\ket{\Phi}\in\Hmc$ that $\bra{\Phi}\Pneq\ket{\Phi}$ is negligibly small, then our observer almost certainly concludes that the state $\ket{\Phi}$ is in thermal equilibrium.
In such a situation, we can simply say that the state $\ket{\Phi}$ is in thermal equilibrium.

We stress that the microcanonical energy shell $\Hmc$ contains plenty of highly nonequilibrium states, in which the probability $\bra{\Phi}\Pneq\ket{\Phi}$ is not small, or even one.\footnote{%
Nevertheless, an overwhelming majority of states in $\Hmc$ are in thermal equilibrium.
This fact is known as the typicality of thermal equilibrium.
}
See section~\ref{s:noneq}.
This fact is crucial for us since we need to take nonequilibrium initial states to discuss thermalization.

We note in passing that it is a common misconception, probably fostered by traditional-style textbooks in statistical mechanics, that the prediction of equilibrium statistical mechanics should always be compared with an averaged quantity in the corresponding physical system.
As is clear from the above discussion (and from our main result to be discussed below), it may happen that the outcomes of a single simultaneous measurement coincide with the statistical mechanical equilibrium values within a high precision, provided that both the system and the quantity to be measured are macroscopic.
One may say that this is a manifestation of the law of large numbers.

\subsection{Main theorems}
\label{s:maintheorems}
In what follows, we fix constants $L$, $\epo>0$, $\eta>0$, $u_0\in(0,\epo)$, $\Duo$, $m$, and $\Du$.
As we already discussed, we assume $\eta$, $\Duo/u_0$, and $\Duo/\Du$ are small.
We understand, however, that these quantities are of ``order one'' in the sense that they are independent of the system size $L$, which we assume to be macroscopically large.
In addition, we assume that $\eta^{-1}$ is an integer and $2L+1$ is a prime.
These somewhat exotic assumptions are necessary to prove the absence of degeneracy in the energy spectrum.
See Lemma~\ref{L:nondeg}.
Although the above assumptions are enough to prove our theorems, we shall make the following (rather technical) assumptions to make various estimates quantitative:\footnote{%
To prove our theorems, we can replace the assumption \rlb{Du0ep} by a weaker assumption $\eta\le1/50$, which is \rlb{etasmall}.
We assumed \rlb{Du0ep} to guarantee that the energy shell $\Hmc$ contains nonequilibrium states.
See section~\ref{s:noneq}.
}
\eqg
L\ge\frac{3m\epo}{\eta u_0},
\lb{L>}\\
\frac{\Duo}{u_0}\le\frac{1}{50},
\lb{u0small}\\
22\Duo\le\Du\le\frac{u_0}{2},
\lb{Duconditions}\\
\Duo\ge\eta\epo.
\lb{Du0ep}
\eng

For a given initial state $\kPz$, we denote its time evolution by
\eq
\kPt=e^{-i\hH t}\kPz.
\en
Let us first state the ``ergodic version'' of our thermalization theorem.
The theorem is proved in section~\ref{s:proof1}, assuming two basic lemmas.

\begin{theorem}[ergodicity theorem]\label{T:main1}
Let $\kPz$ be an arbitrary normalized state from the microcanonical energy shell $\Hmc$ defined in \rlb{Hmc}.
Then one has
\eq
\lim_{T\up\infty}\frac{1}{T}\int_0^Tdt\bPt\Pneq\kPt\le C_1\exp\Bigl[-\frac{(\Du)^2}{8(m-1)\epo u_0}\,L\Bigr]
\lb{ergbound}
\en
with 
\eq
C_1=m^{\frac{2m}{\eta}+\frac{3}{2}}+m\Bigl(\frac{m}{m-1}\Bigr)^{\frac{2m}{\eta}+\frac{1}{2}}.
\lb{C1C2}
\en
\end{theorem}

We have not tried to optimize the constants in \rlb{ergbound}.
One might notice that the constant $C_1$ diverges as $\eta\dn0$, but there is nothing essential about this behavior.
In fact, this constant is originally written as
\eq
C_1=m^{\frac{2}{\delta}+\frac{3}{2}}+m\Bigl(\frac{m}{m-1}\Bigr)^{\frac{2}{\delta}+\frac{1}{2}}.
\lb{C1org}
\en
with another independent small constant $\delta>0$.
We here set $\delta=\eta/m$ for notational simplicity.
See section~\ref{s:proof_ETH_2}.

Consider the natural setting where $\frac{(\Du)^2}{\epo u_0}$ is of order one and $L$ is macroscopically large so that $\frac{(\Du)^2}{\epo u_0}L\gg1$.
Then, the right-hand side of \rlb{ergbound} becomes negligibly small.  
We see that the long-time average of the expectation value of the nonequilibrium projection $\Pneq$ is essentially zero.

This observation suggests that the system is mostly in thermal equilibrium in the long run.
In fact, one can easily convert Theorem~\ref{T:main1} into the following statement relevant to an instantaneous measurement of the energy densities.
See the end of the present subsection for the proof.
Here we take a constant $C_2>0$ such that $(C_2)^2>C_1$.

\begin{theorem}[thermalization at large and typical time]\label{T:main2}
Let $\kPz$ be an arbitrary normalized state from $\Hmc$.
For any sufficiently large $T>0$ (where how large $T$ should be may depend on $\kPz$) there exists a subset of ``atypical moments'' $\calA\subset[0,T]$ such that one has
\eq
\bPt\Pneq\kPt
\le C_2\exp\Bigl[-\frac{(\Du)^2}{16(m-1)\epo u_0}\,L\Bigr],
\lb{mainbound}
\en
for any $t\in[0,T]\backslash\calA$.
If we denote by $\ell(\calA)$ the total length (or the Lebesgue measure) of $\calA$, we have
\eq
\frac{\ell(\calA)}{T}\le C_2\exp\Bigl[-\frac{(\Du)^2}{16(m-1)\epo u_0}\,L\Bigr].
\lb{ellA}
\en
\end{theorem}

Again, assume that $\frac{(\Du)^2}{\epo u_0}L\gg1$, which means that the right-hand sides of \rlb{mainbound} and \rlb{ellA} are negligibly small.
From \rlb{ellA}, we see that the subset $\calA$ is extremely minor in the whole interval $[0,T]$.  We can say that it is atypical for a moment $t\in[0,T]$ to belong to $\calA$.
Take any typical moment, i.e., $t\in[0,T]\backslash\calA$ and consider the corresponding time-evolved state $\kPt$.
Recalling the definition \rlb{Pneq} of $\Pneq$, we see from \rlb{mainbound} that if the observer makes a simultaneous measurement of the energy densities $\hu_1,\ldots,\hu_m$, then all the measurement result, with probability almost equal to one, coincide with the equilibrium value $u_0$ within the precision $\Du$.

Informally speaking, Theorem~\ref{T:main2} establishes that, for a sufficiently large and typical time $t$, the measurement outcomes of the energy densities in the $m$ pieces in the time-evolved state $\ket{\Phi(t)}$ are almost certainly equal to $u_0$.
This means that the state $\kPt$ is in thermal equilibrium according to our criterion based on the measurement of the coarse-grained energy distribution.
It is essential here that we are dealing with the result of a single quantum mechanical simultaneous measurement rather than quantum mechanical averages.
Recall that the latter are obtained through repeated measurements in an ensemble of states.

Suppose that the initial state $\kPz$ is a nonequilibrium state in the sense that $\bPz\Pneq\kPz$ is not small.
We shall see in section~\ref{s:noneq} that the microcanonical energy shell $\Hmc$ contains plenty of nonequilibrium states.
Then Theorem~\ref{T:main2} precisely states that the nonequilibrium initial state relaxes to thermal equilibrium after a sufficiently long and typical time.
Unfortunately, we (still) do not have any quantitative estimates of the relaxation time.
If the initial state $\kPz$ is in thermal equilibrium, i.e., $\bPz\Pneq\kPz$ is small from the beginning, then Theorem~\ref{T:main2} shows the stability of thermal equilibrium.

\medskip

\noindent
{\em Proof of Theorem~\ref{T:main2} assuming Theorem~\ref{T:main1}}\/:
We write
\eq
\calW=\frac{(\Du)^2}{16(m-1)\epo u_0}\,L,
\en
so that \rlb{ergbound} reads
\eq
\lim_{T\up\infty}\frac{1}{T}\int_0^Tdt\bPt\Pneq\kPt\le C_1\,e^{-2\calW}.
\en
Since $C_1<(C_2)^2$, there exists $T_0>0$ such that one has 
\eq
\frac{1}{T}\int_0^Tdt\bPt\Pneq\kPt\le (C_2\,e^{-\calW})^2,
\lb{ergboundT}
\en
for any $T\ge T_0$.
Fix an arbitrary $T$ such that $T\ge T_0$ and define
\eq
\calA=\bigl\{t\in[0,T]\,\bigr|\,\bPt\Pneq\kPt>C_2\,e^{-\calW}\bigr\}.
\en
The first statement of the theorem follows from this definition.
We next observe that
\eq
\frac{\ell(\calA)}{T}=\frac{1}{T}\int_0^Tdt\,\theta\Bigl(\frac{\bPt\Pneq\kPt}{C_2\,e^{-\calW}}-1\Bigr),
\en
where the step function is defined by $\theta(x)=1$ for $x>0$ and $\theta(x)=0$ for $x\le0$.
Noting that $\theta(x-1)\le x$, we find
\eq
\frac{\ell(\calA)}{T}\le\frac{1}{T}\int_0^Tdt\,\frac{\bPt\Pneq\kPt}{C_2\,e^{-\calW}}\le C_2\,e^{-\calW},
\en
where we used \rlb{ergboundT}.
This proves the second statement of the theorem.~\qedm

\subsection{Basic Lemmas and the proof of Theorem~\ref{T:main1}}
\label{s:proof1}
In this section, we prove the ergodicity theorem, Theorem~\ref{T:main1}, by assuming two Lemmas.
These lemmas are essential ingredients of the present theory of thermalization. 
Their proofs are based on the exact mapping of our model of one-dimensional solid to a free fermion model.

The first lemma, Lemma~\ref{L:nondeg} below, guarantees that the energy eigenvalues of our Hamiltonian \rlb{HS} is free from degeneracy.
In general, it is believed that the energy eigenvalues of a quantum many-body system are nondegenerate unless there are special reasons, such as symmetry or integrability, that cause degeneracy.
Even when there are accidental degeneracies, they may always be lifted by adding an appropriate small perturbation to the Hamiltonian.
It is, however, not at all easy (if not impossible) to make this intuition into proof for a concrete class of models.\footnote{As far as we know, the absence of degeneracy in the energy eigenvalues of non-random quantum many-body systems was proved only for free fermion chains by us in \cite{Tasaki2010,Tasaki2016,ShiraishiTasaki2023}.
}
For the present model, which is indeed integrable, we have the following lemma, which will be proved in section~\ref{s:proof_nondeg}.

\begin{lemma}[nondegeneracy of the energy spectrum]\label{L:nondeg}
Let $2L+1$ be a prime and $\eta^{-1}$ be an integer.
Then, all the energy eigenvalues of the Hamiltonian \rlb{HS} are nondegenerate.
\end{lemma}

Recall that the extra potential at site $L$ in the Hamiltonian \rlb{HS} breaks the reflection symmetry, which is present in the standard open chain and causes degeneracy.
The requirement that $2L+1$ is a prime does not look physical since the properties of a physical system on the long chain should not depend on whether the chain length is prime or not.
Nevertheless, we do need this condition to rule out degeneracy completely in our system.

The second lemma, Lemma~\ref{L:ETH} below, establishes a version of strong ETH (energy eigenstate thermalization hypothesis).
In general, a strong ETH is a statement that every energy eigenstate (in the microcanonical energy shell) is in thermal equilibrium according to a certain criterion \cite{vonNeumann,GLTZ,GLMTZ09b,Tasaki2010,Tasaki2016,Deutsch1991,Srednicki1994,Tasaki1998,RigolSrednicki2012,DAlessioKafriPolkovnikovRigol2016}.
For the present model, we prove the following lemma in sections~\ref{s:proof_ETH_1} and \ref{s:proof_ETH_2}.

\begin{lemma}[large-devitaion type ETH]\label{L:ETH}
Assume that the conditions for the constants stated at the beginning of section~\ref {s:maintheorems} are valid.
Then, for any normalized eigenstate $\kPsi$ of the Hamiltonian \rlb{HS} that belongs to the microcanonical energy shell $\Hmc$, one has
\eq
\bPsi\Pneq\kPsi\le C_1\exp\Bigl[-\frac{(\Du)^2}{8(m-1)\epo u_0}\,L\Bigr],
\lb{ETH}
\en
where the right-hand side is exactly equal to that in \rlb{ergbound}.
\end{lemma}

When $\frac{(\Du)^2}{\epo u_0}$ is of order one and $L$ is macroscopically large so that $\frac{(\Du)^2}{\epo u_0}L\gg1$, \rlb{ETH} implies that the energy eigenstate $\kPsi$ itself is in thermal equilibrium according to our criterion.
This is nothing but the strong ETH for the energy densities $\hu_1,\ldots,\hu_m$ in the large-deviation form formulated by us in \cite{Tasaki2016}.

Assuming these two lemmas, Theorem~\ref{T:main1} is easily proved as follows.
We note that the following proof is general and does not rely on the specific characters of the present model.
This means that Theorems~\ref{T:main1} and \ref{T:main2} can be proved for any quantum many-body systems provided that one has statements corresponding to Lemmas~\ref{L:nondeg} and \ref{L:ETH}.

\medskip\noindent
{\em Proof of Theorem~\ref{T:main1} assuming Lemmas~\ref{L:nondeg} and \ref{L:ETH}}\/:
The proof is standard and easy.
Let $\ket{\Psi_j}$ be a normalized eigenstate of $\hH$ with energy eigenvalue $E_j$.
We relabel the index so that $\ket{\Psi_j}$ with $j=1,\ldots,D$ precisely span the microcanonical energy shell $\Hmc$.
Expanding the initial state as
\eq
\kPz=\sum_{j=1}^D\alpha_j\ket{\Psi_j},
\en 
we see that the time-evolved state is explicitly given by
\eq
\kPt=\sum_{j=1}^De^{-iE_jt}\,\alpha_j\ket{\Psi_j}.
\en
Then, the expectation value in the integrand in the left-hand side of \rlb{ergbound} is written as
\eq
\bra{\Phi(t)}\Pneq\ket{\Phi(t)}=\sum_{j,j'=1}^De^{i(E_j-E_{j'})t}\,\alpha_j^*\alpha_{j'}\bra{\Psi_j}\Pneq\ket{\Psi_{j'}}.
\lb{PPP}
\en
Since Lemma~\ref{L:nondeg} guarantees $E_j\ne E_{j'}$ whenever $j\ne j'$, the long-time average of \rlb{PPP} becomes
\eq
\lim_{T\up\infty}\frac{1}{T}\int_0^Tdt\,\bra{\Phi(t)}\Pneq\ket{\Phi(t)}=\sum_{j=1}^D|\alpha_j|^2\bra{\Psi_j}\Pneq\ket{\Psi_j}.
\en
We then get the desired \rlb{ergbound} from \rlb{ETH}.~\qedm

\subsection{Why do we have thermalization in an integrable model?}
\label{s:integrable}
It is often said that an integrable quantum many-body system does not exhibit thermalization.
The reader might be then puzzled to learn that our model of solid, which exhibits thermalization, is equivalent to a typical integrable model, namely, the free fermion chain.
Let us clarify the issue.

The main reason that integrable models are said not to thermalize is that they generally possess many conserved quantities.
Suppose $\hA$ is conserved (in the sense that $[\hH,\hA]=0$) and choose a nonequilibrium initial state $\kPz\in\Hmc$ such that $\bPz\hA\kPz\ne\sbkt{\hA}_\mathrm{mc}$.
Since $\bra{\Phi(t)}\hA\ket{\Phi(t)}$ is independent of $t$, the state $\ket{\Phi(t)}$ is never in thermal equilibrium if one uses the expectation value of $\hA$ as a criterion for thermal equilibrium.

Our model also possesses a number of conserved quantities.
We nevertheless observe (and can prove the presence of) thermalization because we use the phenomenological and operational characterization of thermal equilibrium discussed in section~\ref{s:criterion}.
If our observer can measure quantities other than the coarse-grained energy distribution and decides to use them also to judge if a state is in thermal equilibrium, then it is possible that she/he concludes that our system does not thermalize.\footnote{%
A trivial example is given by the observable $\hA=\sum_{x=1}^{L-1}(\had_x\ha_{x+1}+\had_{x+1}\ha_x)-\hn_L$, which is nothing but the hopping part of the Hamiltonian \rlb{HS}. It obviously commutes with $\hH$.
By using the mapping to a free fermion chain (see section~\ref{s:freefermion}), one can show that the thermal expectation value $\sbkt{\hA}_\mathrm{mc}$ is at most $O(1)$, while it is possible to prepare an initial state $\ket{\Phi(0)}\in\Hmc$ that is an eigenstate of $\hA$ with the eigenvalue of $O(L)$.
The state $\ket{\Phi(0)}$ never thermalizes if one includes the observable $\hA$ in the criterion of thermal equilibrium.
%
}
(We note in passing that it is physically reasonable to concentrate only on the coarse-grained energy distribution if one is interested in thermalization in a weakly heat-conducting solid.)
If we use the language from \cite{GoldsteinHuseLebowitzTumulka2015}, our system approaches thermal equilibrium in the sense of MATE (macroscopic thermal equilibrium) but not in the sense of MITE (microscopic thermal equilibrium).

%
%
%

We also note that thermalization, or, more generally, equilibration, in free fermion systems has been established theoretically \cite{GluzaEisertFarrelly2019,ShiraishiTasaki2023,TasakiFreeFermion} and numerically \cite{RigolMuramatsuOlshanii2006,RigolFitzpatrick2011,Pandeyetal} in various settings.

\section{Proofs of Lemmas~\ref{L:nondeg} and \ref{L:ETH}}
\label{s:proof}
In this section, which is a technical core of the present note, we prove two basic lemmas.
We first observe that the model is equivalent to a free fermion chain and write down exact energy eigenstates and eigenvalues (section~\ref{s:freefermion}).
As a simple application, we show that the microcanonical energy shell $\Hmc$ contains many equilibrium states (section~\ref{s:noneq}).
Then we prove Lemma~\ref{L:nondeg} for nondegeneracy of energy eigenvalues by using a classical number theoretic theorem (section~\ref{s:proof_nondeg}).
We finally prove Lemma~\ref{L:ETH} that establishes the strong ETH bound (sections~\ref{s:proof_ETH_1} and \ref{s:proof_ETH_2}).

\subsection{Free fermion chain}
\label{s:freefermion}
A key for the proofs of the two basic lemmas is that our model of a one-dimensional ``solid'' with the Hamiltonian \rlb{HS} is equivalent to a model of spinless free fermions.

Let us first define a free fermion system on the chain $\{1,\ldots,L\}$ precisely.
For each $x\in\{1,\ldots,L\}$, let $\hc_x$ and $\hcd_x$ be the annihilation and the creation operators, respectively, of a fermion at site $x$.
They satisfy the standard anticommutation relations\footnote{
We write $\{\hA,\hB\}=\hA\hB+\hB\hA$.
}
\eq
\{\hcd_x,\hc_{y}\}=\delta_{x,y},\quad
\{\hc_x,\hc_{y}\}=\{\hcd_x,\hcd_{y}\}=0,
\lb{AC}
\en
for any $x,y\in\{1,\ldots,L\}$.
We define the number operator at site $x$ by $\hn_x=\hcd_x\hc_x$.
We consider states with all possible fermion numbers.
The whole Hilbert space is spanned by the states of the form $\hcd_{x_1}\ldots\hcd_{x_N}\kv$, where $x_i\in\{1,\ldots,L\}$ with $x_i<x_{i+1}$ and $N\in\{0,1,\ldots,L\}$.
Here $\kv$ is the normalized state with no fermions in the system, which satisfies $\hc_x\kv=0$ for any $x$.

Let us relate the basis states \rlb{basis} of the model of solid and the basis states of the fermion model by a one-to-one correspondence as
\eq
\ket{\mu_1,\mu_2,\ldots,\mu_L}\ \longleftrightarrow\ (\hcd_1)^{\mu_1}(\hcd_2)^{\mu_2}\cdots(\hcd_L)^{\mu_L}\kv,
\lb{basiscor}
\en
where we define $(\hcd_x)^0=1$.
It is easily found that the action of the Hamiltonian \rlb{HS} on the basis $\{\ket{\mu_1,\ldots,\mu_L}\}$ is precisely recovered by the free fermion Hamiltonian
\eq
\hH=\epo\sum_{x=1}^L\hn_x+\frac{\eta\epo}{2}\Bigl\{\sum_{x=1}^{L-1}(\hcd_x\hc_{x+1}+\hcd_{i+1}\hc_x)-\hn_L\Bigr\},
\lb{H}
\en
acting on the standard free fermion basis \rlb{basiscor}.
(Here, we make a slight abuse of notation and use the same symbols $\hH$ and $\hn_x$ as before to mean the operators of the fermion system.)
Recall that such a naive correspondence between a bosonic model (like our model of solid) and a fermion model breaks down when a successive action of the Hamiltonian exchanges particles.
We are free from this problem since no exchange can occur in our model on the open chain with nearest neighbor hopping.
It should also be stressed that the introduction of the fermion model has no physical significance.
It is only a mathematical tool for conveniently expressing antisymmetrization involved in the expressions of the energy eigenstates.

\begin{figure}
\centerline{\epsfig{file=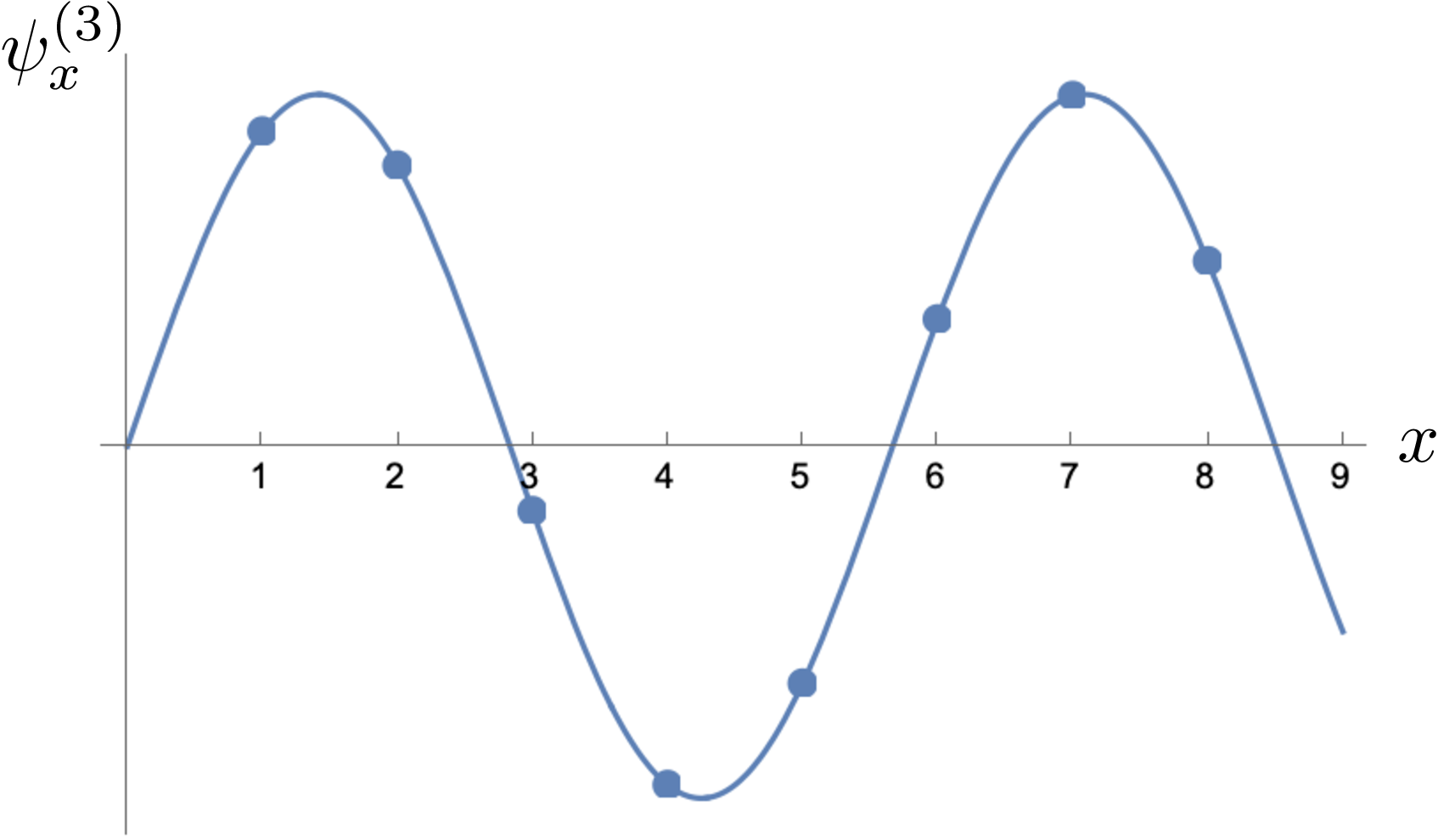,width=8truecm}}
\caption[dummy]{
The single particle energy eigenstate $\psi^{(3)}_x$ for the model with $L=8$.
Note that $\psi^{(3)}_0=0$ and $\psi^{(3)}_8=-\psi^{(3)}_9$, and hence the boundary equations \rlb{Sch2} and \rlb{Sch3} are satisfied.
}
\label{f:sin}
\end{figure}

It is standard and elementary to write down the energy eigenstates and eigenvalues of the Hamiltonian \rlb{H}.
The single-particle Schr\"odinger equation corresponding to \rlb{H} is
\eqa
&\ep\,\psi_x=\epo\,\psi_x+\frac{\eta\epo}{2}(\psi_{x-1}+\psi_{x+1}),\quad x=2,\ldots,L-1;\lb{Sch1}\\
&\ep\,\psi_1=\epo\,\psi_1+\frac{\eta\epo}{2}\psi_2;\lb{Sch2}\\
&\ep\,\psi_L=\epo\,\psi_L+\frac{\eta\epo}{2}(\psi_{L-1}-\psi_L),\lb{Sch3}
\ena
where $\ep$ is the single-particle energy eigenvalue.
From \rlb{Sch1}, one readily sees that the energy eigenstate has the form $\psi_x=A\sin(kx)$, where $k>0$ is the wave number to be determined, and the corresponding energy eigenvalue is $\ep=\epo+\eta\epo\cos k$.
Note that \rlb{Sch2} is satisfied because $\psi_0=0$.
By comparing \rlb{Sch3} with \rlb{Sch1}, we see that it is necessary that $\psi_{L+1}=-\psi_L$ for the above wave function to become an eigenstate.
An inspection shows that it suffices to assume $\psi_{L+(1/2)}=0$, i.e.,
\eq
\sin\Bigl(k\Bigl(L+\frac{1}{2}\Bigr)\Bigr)=0.
\en
We thus find the wave numbers should be
\eq
k=\frac{\pi j}{\Lh},\quad j=1,\ldots,L.
\en
For $j=1,\ldots,L$, the normalization single-particle energy eigenstate is
\eq
\psi^{(j)}_x=\sqrt{\frac{2}{\Lh}}\sin\Bigl(\frac{\pi j}{\Lh}x\Bigr),
\lb{psijx}
\en
and the corresponding single-particle energy eigenvalue is
\eq
\ep_j=\epo+\eta\epo\cos\Bigl(\frac{\pi j}{\Lh}\Bigr).
\lb{ej}
\en
This solves the single-particle eigenvalue problem since we have obtained $L$ eigenfunctions with distinct eigenvalues.
We used the general formula
\eq
\sum_{x=\ell_1}^{\ell_2}\{\sin(kx)\}^2=\frac{1}{4}\Bigl\{2(\ell_2-\ell_1+1)+\frac{\sin(k(2\ell_1-1))}{\sin k}-\frac{\sin(k(2\ell_2+1))}{\sin k}\Bigr\},
\lb{sinsum}
\en
with $\ell_1=1$ and $\ell_2=L$
to determine the normalization constant in \rlb{psijx}.

To discuss the corresponding many-particle problem, we introduce the fermion operator
\eq
\hb_j=\sum_{x=1}^L\psi^{(j)}_x\,\hc_x,
\lb{bj}
\en
for $j=1,\ldots,L$.
One finds from \rlb{AC} and the orthonormality of the above set of energy eigenstates that these operators satisfy the standard anticommutation relations
\eq
\{\hbd_j,\hb_{j'}\}=\delta_{j,j'},\quad\{\hb_j,\hb_{j'}\}=\{\hbd_j,\hbd_{j'}\}=0.
\en
The Hamiltonian \rlb{H} is diagonalized as
\eq
\hH=\sum_{j=1}^L\ep_j\,\hbd_j\hb_j.
\en
The corresponding many-particle energy eigenstate and eigenvalues are specified by a sequence $(j_1,\ldots,j_N)$, where $j_i\in\{1,\ldots,L\}$ with $j_i<j_{i+1}$ and $N\in\{0,1,\ldots,L\}$.
The normalized energy eigenstate is
\eq
\ket{\Psi_{(j_1,\ldots,j_N)}}=\hbd_{j_1}\cdots\hbd_{j_N}\kv,
\lb{Psijj}
\en
and the corresponding energy eigenvalue is
\eq
E_{(j_1,\ldots,j_N)}=\sum_{i=1}^N\ep_{j_i}.
\lb{Ejj}
\en

\subsection{The existence of nonequilibrium initial states}
\label{s:noneq}
In this short subsection, we prove that the microcanonical energy shell $\Hmc$ contains many nonequilibrium states.
Assume, for simplicity, that there is an integer $N$ such that $u_0=N\epo/L$.
From \rlb{ej} and \rlb{Ejj}, we see that any energy eigenvalue with $N$ fermions satisfy
\eq
N(\epo-\eta\epo)\le E\le N(\epo+\eta\epo),
\en
which means
\eq
\Bigl|\frac{E}{L}-u_0\Bigr|\le\eta\epo\frac{N}{L}\le\eta\epo\le\Duo,
\en
where we used the assumptiion \rlb{Du0ep}.
Recalling the definition \rlb{Hmc} of the energy shell, we see that any $N$ fermion state is a member of  $\Hmc$.

Then, it is apparent that $\Hmc$ contains many nonequilibrium states.
For example, for any sequence $x_1,\ldots,x_N\in\{1,\ldots,L\}$ such that $x_i<x_{i+1}$ the corresponding $N$ fermion state $\hcd_{x_1}\ldots\hcd_{x_N}\ket{\Phi_{\rm vac}}$ is in $\Hmc$.
The state may be in thermal equilibrium or not depending on the configuration $x_1,\ldots,x_N$.
We can choose a drastic nonequilibrium state in which, for example, all the fermions, i.e., excited atoms, are densely packed in one end of the solid.

\subsection{Proof of Lemma~\ref{L:nondeg}}
\label{s:proof_nondeg}
We shall prove that the energy eigenvalues \rlb{Ejj} are free from degeneracy.
The proof makes use of a particular form of the energy eigenvalues of the free fermion model and a theorem from number theory.
The strategy of the proof, therefore, works only for some free fermion chains.\footnote{%
As far as we know, the proof of the absence of degeneracy in certain free fermion chains based on number theoretic results first appeared in \cite{Tasaki2010,Tasaki2016}.
A stronger result was proved in \cite{ShiraishiTasaki2023} by using a new number theoretic theorem.
The new result summarized in Lemma~\ref{L:nondeg} is more satisfactory than the previous results since it does not require a fine-tuning of the model with a flux parameter. 
}

We start by stating an important number theoretic theorem without proof.
The interested reader is suggested to study my lecture video \cite{Hal_NT} for elementary proofs of the theorems.
Let $p$ be an odd prime and 
\eq
\zeta=e^{i 2\pi/p},
\lb{zeta}
\en
be the $p$-th root of unity.

\begin{figure}
\centerline{\epsfig{file=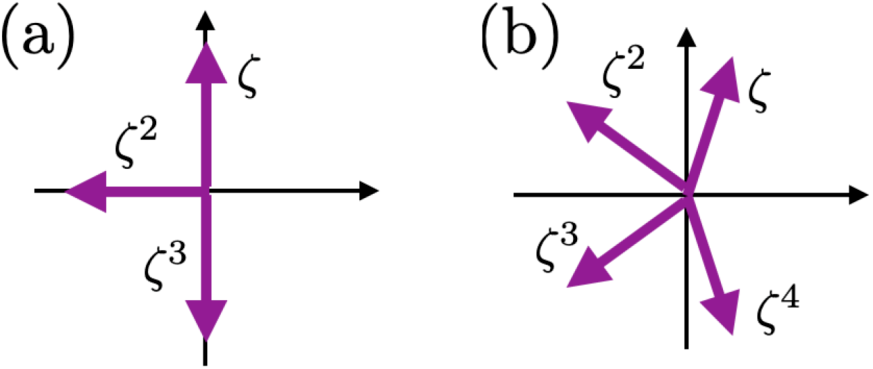,width=8truecm}}
\caption[dummy]{
We show $\zeta,\ldots,\zeta^{p-1}$ as vectors in the complex plane for the cases with (a)~$p=4$ and (b)~$p=5$.
(a)~For $p= 4$, one has $\zeta+\zeta^3=0$ and \rlb{summmune0} is invalid.
(b)~Since $p=5$ is a prime, Theorem~\ref{T:Gauss} implies that $\alpha_1\zeta+\alpha_2\zeta^2+\alpha_3\zeta^3+\alpha_4\zeta^4\ne0$ for any $\alpha_1,\alpha_2,\alpha_3,\alpha_4\in\bbQ$ unless $\alpha_1=\alpha_2=\alpha_3=\alpha_4=0$.
The same conclusion is never valid if $\alpha_1,\alpha_2,\alpha_3,\alpha_4\in\bbR$.
}
\label{f:zeta}
\end{figure}

\begin{theorem}\label{T:Gauss}
For any $m_1,\ldots,m_{p-1}\in\bbZ$ such that $m_\mu\ne0$ for some $\mu$, one has
\eq
\summu m_\mu\,\zeta^\mu\ne0.
\lb{summmune0}
\en
\end{theorem}

Here, it is crucial that the sum is from 1 to $p-1$, rather than from 1 to $p$.
Otherwise \rlb{summmune0} can never be true because $\sum_{\mu=1}^p\zeta^\mu=0$ for any $p$.
The assumption that $p$ is a prime is essential.
When $p=4$, for example, one has $\zeta+\zeta^3=1$, violating \rlb{summmune0}.
See Figure~\ref{f:zeta}~(a).
It is also essential that $m_\mu$ are integers.
Note that \rlb{summmune0} says that the two-dimensional vectors $\zeta,\ldots,\zeta^{p-1}$ in the complex plane are linearly independent, provided that the coefficients are integers.  (It is easy to observe that the same conclusion holds for rational coefficients.)
However, in standard linear algebra with real coefficients, at most two vectors can be linearly independent.
See Figure~\ref{f:zeta}~(b).

The theorem is a straightforward consequence of the classical result by Gauss, known as the irreducibility of the cyclotomic polynomials of prime index.
The proof can be found in standard texts in number theory (as well as in my video  \cite{Hal_NT}).
See, e.g., Chapter 12, Section 2 of \cite {Tignol}, and also Chapter 13, Section 2 of \cite{IR} or section~3.2 of \cite{StewarTall}.

To prove Lemma~\ref{L:nondeg}, it is convenient to introduce the standard occupation number description.
For a given sequence $(j_1,\ldots,j_N)$, we define the corresponding occupation numbers $(n_1,\ldots,n_L)$ as
\eq
n_j=\begin{cases}
1,&\text{if $j=j_i$ for some $i$};\\
0,&\text{otherwise}.
\end{cases}
\en
By using the occupation numbers, the energy eigenvalue \rlb{Ejj} with \rlb{ej} is written as
\eq
E_{(n_1,\ldots,n_L)}=\sum_{j=1}^Ln_j\ep_j
=\epo\,N+\eta\epo\sum_{j=1}^Ln_j\cos\Bigl(\frac{\pi j}{\Lh}\Bigr),
\lb{En}
\en
where $N=\sum_{j=1}^Ln_j$ is the total number of fermions.
Let us write $p=2L+1$, which is a prime by our assumption.
Noting that
\eq
\cos\Bigl(\frac{\pi j}{\Lh}\Bigr)=\frac{1}{2}\bigl\{e^{i2\pi j/p}+e^{-i2\pi j/p}\bigr\},
\en
we can express the energy eigenvalue as
\eqa
E_{(n_1,\ldots,n_L)}&=\frac{\eta\epo}{2}\Bigl(KN+\sum_{j=1}^{(p-1)/2}n_j(\zeta^j+\zeta^{p-j})\Bigr)
\nl&=\frac{\eta\epo}{2}\Bigl(KN+\sum_{\mu=1}^{(p-1)/2}n_\mu\,\zeta^\mu+\sum_{\mu=(p+1)/2}^{p-1}n_{p-\mu}\,\zeta^\mu\Bigr),
\lb{Enrep}
\ena
where $K=2/\eta$.
Recall that $K$ is an integer by our assumption about $\eta$.

Take arbitrary two distinct energy eigenstates labeled by $(n_1,\ldots,n_L)$ and  $(n'_1,\ldots,n'_L)$.
Let us also write $\Di N=\sum_{j=1}^Ln_j-\sum_{j=1}^Ln'_j$, and assume $\Di N\ge0$ without losing generality.
By using \rlb{Enrep}, we see
\eqa
\frac{2}{\eta\epo}\bigl(E_{(n_1,\ldots,n_L)}-E_{(n'_1,\ldots,n'_L)}\Bigr)&=K\Di N+\sum_{\mu=1}^{(p-1)/2}(n_\mu-n'_\mu)\,\zeta^\mu+\sum_{\mu=(p+1)/2}^{p-1}(n_{p-\mu}-n'_{p-\mu})\,\zeta^\mu
\nl&=\summu m_\mu\,\zeta^\mu,
\ena
with
\eq
m_\mu=\begin{cases}
n_\mu-n'_\mu-K\Di N,&\mu=1,\ldots,(p-1)/2;\\
n_{p-\mu}-n'_{p-\mu}-K\Di N,&\mu=(p+1)/2,\ldots,p-1.
\end{cases}
\en
Here we noted $\sum_{\mu=1}^{p-1}\zeta^\mu=-1$.

We now claim that $m_\mu\ne0$ for some $\mu$.
We then see from Theorem~\ref{T:Gauss} that $E_{(n_1,\ldots,n_L)}\ne E_{(n'_1,\ldots,n'_L)}$, which proves Lemma~\ref{L:nondeg}.

To show the claim, assume $m_\mu=0$ for all $\mu$.
This implies $n_j-n'_j=K\Di N\ge0$ for all $j=1,\ldots,L$.
Since $(n_1,\ldots,n_L)\ne(n'_1,\ldots,n'_L)$, the difference $n_j-n'_j$ can be independent of $j$ only when $n_j=1$ and $n'_j=0$ for all $j=1,\ldots,L$.
But this implies $1=K\Di N$ with $\Di N=L$, which is impossible as $K$ is a (large) integer.

\subsection{The essence of the proof of Lemma~\ref{L:ETH}}
\label{s:proof_ETH_1}
Recall that the normalized single-particle energy eigenstates are given in \rlb{psijx}.
For each $q=1,\ldots,m$ and $j=1,\ldots,L$, we define
\eq
p_q^{(j)}=\sum_{x\in\La_q}\bigl|\psi^{(j)}_x\bigr|^2,
\lb{pLdef}
\en
which is the probability of finding a particle in the sublattice $\La_q$ defined in \rlb{Laq} in the $j$-the single-particle energy eigenstate.
We also denote the number of fermions in $\La_q$ as
\eq
\hN_q=\sum_{x\in\La_q}\hn_x.
\lb{NLdef}
\en
The following simple lemma proved at the end of this section is the technical key of the present work.
It was proved in our previous work \cite{TasakiFreeFermion} by using the idea in \cite{ShiraishiTasaki2023}.

\begin{lemma}\label{L:eN}
For any normalized many-particle energy eigenstate \rlb{Psijj}, any $q=1,\ldots,m$, and any $\la\in\bbR$, we have
\eq
\bra{\Psi_{(j_1,\ldots,j_N)}}e^{\la\hN_q}\ket{\Psi_{(j_1,\ldots,j_N)}}\le\prod_{i=1}^N\bigl\{e^\la\,p_q^{(j_i)}+(1-p_q^{(j_i)})\bigr\}.
\lb{PelaNP}
\en
\end{lemma}

In what follows, we heuristically describe the central idea of the proof of Lemma~\ref{L:ETH}, making some uncontrolled approximations.
For simplicity, we shall focus on the case with $m=2$.
A full proof, which is rather technical and may be skipped, is given in section~\ref{s:proof_ETH_2}.
We now take an arbitrary energy eigenstate $\ket{\Psi_{(j_1,\ldots,j_N)}}$ that belongs to $\Hmc$, and abbreviate it as $\kPsi$.

First, it seems reasonable to assume $p_q^{(j)}\simeq1/m=1/2$ for all $j$ since the energy-eigenstate wave functions should spread almost uniformly over the chain.
(In fact, this is not the case for some exceptional states, and we must control the deviation carefully.
See Lemma~\ref{L:p<} below.)
We then have
\eq
\bPsi e^{\la\hN_q}\kPsi\lesssim\Bigl(\frac{e^\la}{2}+\frac{1}{2}\Bigr)^N
=\Bigl(\cosh\frac{\la}{2}\Bigr)^N\,e^{\la N/2}.
\lb{PePrough}
\en

Secondly, we shall approximate $u_0L\simeq \epo N$ and $\hH_q\simeq\epo\hN_q$ since $\eta$ is small.
Then the condition $\hu_q-u_0=\frac{2}{L}\hH_q-u_0\ge\Du$ is rewritten as $\hN_q\gtrsim(N+\DN)/2$ with $\DN=L\Du/\epo$.
Therefore, we see for any $\la\ge0$ that\footnote{%
Let $\hA$ be a self-adjoint operator and $\ket{\Xi}$ be its eigenstate, i.e., $\hA\ket{\Xi}=a\ket{\Xi}$.
We define  $\Proj[\hA\ge a_0]\ket{\Xi}=\ket{\Xi}$ if $a\ge a_0$ and $\Proj[\hA\ge a_0]\ket{\Xi}=0$ if $a<a_0$.
}
\eqa
\bPsi\Proj[\tfrac{2}{L}\hH_q-u_0\ge\Du]\kPsi&\simeq
\bPsi\Proj[\hN_q-\tfrac{1}{2}(N+\DN)\ge0]\kPsi
\nl&
\le\bPsi e^{\la\{\hN_q-(N+\DN)/2\}}\kPsi,
\itext{where we used the simple inequality $\Proj[\hA\ge0]\le e^{\hA}$ for a self-adjoint operator $\hA$.
Then by using \rlb{PePrough}, we can further bound this as}
&\lesssim\Bigl(\cosh\frac{\la}{2}\Bigr)^N\,e^{-\la \DN/2}
\simeq \exp\Bigl[N\Bigr(\frac{\la^2}{8}-\la\frac{\DN}{2N}\Bigr)\Bigr],
\itext{where we used \rlb{PePrough} and the approximation $\cosh(\la/2)\simeq 1+\la^2/8\simeq e^{\la^2/8}$.
We now choose $\la=2\DN/N$ to minimize the right-hand side of \rlb{PePrough2} to get}
&\lesssim e^{-(\DN)^2/(2N)}\sim \exp\Bigl[-\frac{(\Du)^2}{2\,\epo u_0}\,L\Bigr].
\lb{PePrough2}
\ena
Since the expectation value $\bPsi\Proj[\tfrac{2}{L}\hH_q-u_0\le-\Du]\kPsi$ can be bounded almost similarly, we recover (although non-rigorously) the desired inequality \rlb{ETH} for $m=2$ with better constants.

\medskip
\noindent{\em Proof of Lemma~\ref{L:eN}}\/:
Fix $q=1,\ldots,m$.
We note that the fermion operators satisfy
\eq
e^{\la\hN_q/2}\,\hcd_x=\begin{cases}
e^{\la/2}\,\hcd_x\,e^{\la\hN_q/2},&x\in\La_q;\\
\hcd_x\,e^{\la\hN_q/2},&x\not\in\La_q,
\end{cases}
\en
and $e^{\la\hN_q/2}\kv=\kv$.
Then, we find from \rlb{Psijj} that
\eq
e^{\la\hN_q/2}\ket{\Psi_{(j_1,\ldots,j_N)}}=\hdd_{j_1}\cdots\hdd_{j_N}\kv,
\en
with
\eq
\hd_j=e^{\la/2}\sum_{x\in\La_q}\psi^{(j)}_x\,\hc_x+\sum_{x\not\in\La_q}\psi^{(j)}_x\,\hc_x.
\en
We then observe that
\eqa
\bra{\Psi_{(j_1,\ldots,j_N)}}e^{\la\hN_q}\ket{\Psi_{(j_1,\ldots,j_N)}}
&=\bv\hd_{j_N}\cdots\hd_{j_2}\hd_{j_1}\hdd_{j_1}\hdd_{j_2}\cdots\hdd_{j_N}\kv
\nl&\le\snorm{\hd_{j_1}\hdd_{j_1}}\,\bv\hd_{j_N}\cdots\hd_{j_2}\hdd_{j_2}\cdots\hdd_{j_N}\kv
\nl&\le\prod_{i=1}^N\snorm{\hd_{j_i}\hdd_{j_i}},
\lb{PelaNP3}
\ena
where we used the basic property $\bra{\Phi}\hA\ket{\Phi}\le\snorm{\hA}\sbkt{\Phi|\Phi}$ of the operator norm repeatedly.

It remains to estimate the norm $\snorm{\hd_j\hdd_j}$.
From \rlb{AC}, we see
\eq
\{\hdd_j,\hd_j\}=e^\la\,p_q^{(j)}+(1-p_q^{(j)}).
\lb{ddd}
\en
Denoting the right-hand side as $\alpha$, we find $(\hd_j\hdd_j)^2=\hd_j\hdd_j\hd_j\hdd_j=\alpha\,\hd_j\hdd_j$.
This implies that the eigenvalues of $\hd_j\hdd_j$ are either 0 or $\alpha$, and hence the norm $\snorm{\hd_j\hdd_j}$ is equal to $\alpha$.
We therefore get \rlb{PelaNP} from  \rlb{PelaNP3} and \rlb{ddd}.~\qedm

\subsection{Details of the proof of Lemma~\ref{L:ETH}}
\label{s:proof_ETH_2}
Let us start with a technical lemma that shows the rough estimate $p_q^{(j)}\simeq1/m$ is justified except for a finite number of $j$.
It is crucial that the number of exceptions, denoted as $M$, is independent of the system size $L$.

\begin{lemma}\label{L:p<}
Take any (small) $\delta>0$, and assume that the system size satisfies
\eq
L\ge\frac{1}{m\delta}.
\lb{L>cond}
\en
Then, one has
\eq
\Bigl|p_q^{(j)}-\frac{1}{m}\Bigr|\le\delta,
\lb{pLj<}
\en
except for $M$ distinct values of $j$, where
\eq
M\le\frac{2}{\delta}+\frac{1}{2}.
\lb{Mcond}
\en
\end{lemma}

\noindent
{\em Proof}\/:
We write the single-particle energy eigenstate \rlb{psijx} as
\eq
\psi^{(j)}_x=\frac{2}{\sqrt{2L+1}}\,\sin(kx),
\lb{psijx2}
\en
with
\eq
k=\frac{2\pi j}{2L+1},
\lb{kandj}
\en
where $j=1,\ldots,L$.
Note that $k$ runs roughly between 0 and $\pi$.
Let
\eq
x_\mathrm{L}=\frac{L}{m}(q-1)+1,\quad x_\mathrm{R}=\frac{L}{m}q,
\en
be the endpoints of $\La_q$.
By using the general formula \rlb{sinsum} for the summation, we can explicitly evaluate the probability \rlb{pLdef} as
\eq
p_q^{(j)}=\frac{4}{2L+1}\sum_{x=x_\mathrm{L}}^{x_\mathrm{R}}(\sin kx)^2
=\frac{1}{2L+1}\Bigl\{2\frac{L}{m}+\frac{\sin(k(2x_\mathrm{L}-1))}{\sin k}-\frac{\sin(k(2x_\mathrm{R}+1))}{\sin k}\Bigr\}.
\lb{pLj1}
\en
Since $\sin k\ne0$, we find
\eq
\Bigl|p_q^{(j)}-\frac{1}{m}\Bigr|\le\frac{1}{(2L+1)m}+\frac{1}{2L+1}\frac{2}{\sin k}.
\lb{pLj<1}
\en
Since the assumption \rlb{L>cond} implies
\eq
\frac{1}{(2L+1)m}\le\frac{\delta}{2},
\en
the right-hand side of \rlb{pLj<1} is bounded from above by $\delta$ if
\eq
\frac{1}{2L+1}\frac{2}{\sin k}\le\frac{\delta}{2}.
\lb{kcond}
\en

The bound \rlb{kcond} apparently does not hold for $k\in(0,\pi)$ close to 0 or $\pi$.
Let $M$ be the number of $k$ of the form \rlb{kandj} that violates \rlb{kcond}.
Since
$\sin k\ge\frac{2}{\pi}k$ and $\sin k\ge\frac{2}{\pi}(\pi-k)$
for $k\in(0,\pi)$, we can upper-bound $M$ by counting the number of $k$ such that
\eq
\frac{2\pi}{(2L+1)k}\ge\delta
\quad\text{or}\quad\frac{2\pi}{(2L+1)(\pi-k)}\ge\delta,
\en
or, equivalently, the number $j\in\{1,\ldots,L\}$ such that
\eq
j\le\delta^{-1}\quad\text{or}\quad L+\tfrac{1}{2}-j\le\delta^{-1}.
\en
We then get \rlb{Mcond} by inspection.~\qedm

\medskip
We fix the microcanonical energy interval $[(u_0-\Duo)L,(u_0+\Duo)L]$ and the corresponding energy shell $\Hmc$ defined as \rlb{Hmc}.
We take an arbitrary energy eigenstate $\ket{\Psi_{(j_1,\ldots,N)}}\in\Hmc$ and abbreviate it as $\kPsi$.

The particle number $N$ in the fermionic description satisfies $\epo N\simeq u_0 L$.
Let us determine the precise range of $N$.
Since the single-particle energy eigenvalue is given by \rlb{ej}, we find that the energy eigenvalue $E$ for energy eigenstates with $N$ particles satisfy
\eq
(\epo-\eta\epo)N\le E\le (\epo+\eta\epo)N.
\en
Thus, we see that the possible particle number $N$ for $E\in[(u_0-\Duo)L,(u_0+\Duo)L]$ is
\eq
\frac{u_0-\Duo}{(1+\eta)\epo}L\le N\le \frac{u_0+\Duo}{(1-\eta)\epo}L.
\lb{Nrange}
\en

For $q=1,\ldots,m$, we define the projection operators
\eq
\hP^+_q=\Proj[\hu_q\ge u_o+\Du],\quad
\hP^-_q=\Proj[\hu_q\le u_o-\Du],
\lb{P+-def}
\en
where the energy density operator $\hu_q$ is defined in \rlb{uq}.
Note that these $2m$ projection operators commute with each other.
In the rest of the present section, we shall prove
\eqg
\bPsi\hP^+_q\kPsi
\le m^{\frac{2m}{\eta}+\frac{1}{2}}\,\exp\Bigl[-\frac{(\Du)^2}{8(m-1)\epo u_0}L\Bigr],
\lb{main01}\\
\bPsi\hP^-_q\kPsi
\le \Bigl(\frac{m}{m-1}\Bigr)^{\frac{2m}{\eta}+\frac{1}{2}}\,\exp\Bigl[-\frac{(\Du)^2}{8(m-1)\epo u_0}L\Bigr],
\lb{main02}
\eng
for any $q=1,\ldots,m$.
Noting that
\eq
\bPsi\Pneq\kPsi\le\sum_{q=1}^m\bigl(\bPsi\hP^+_q\kPsi+\bPsi\hP^-_q\kPsi\bigr),
\lb{PneqP+-}
\en
we get the desired bound \rlb{ETH} with $C_1$ as in \rlb{C1C2}.

Let us see in detail how the expectation value $\bPsi\hP^+_q\kPsi$ is bounded.
Since $(\epo+\eta\epo)\hN_q\ge\hH_q$, we see that the defining condition $\hu_q-u_0=\frac{m}{L}\hH_q-u_0\ge\Du$ for $\hP^+_q$ implies $\hN_q\ge R$ with
\eq
R=\frac{u_0+\Du}{m(1+\eta)\epo}L.
\lb{Rdef}
\en
We thus find
\eq
\bPsi\hP^+_q\kPsi\le\bPsi\Proj[\hN_q-R\ge0]\kPsi
\le\bPsi e^{\la(\hN_q-R)}\kPsi,
\lb{main3}
\en
for any $\la\ge0$.
The expectation value $\bPsi e^{\la\hN_q}\kPsi$ can be evaluated by using Lemmas~\ref{L:eN} and \ref{L:p<}.
(We shall verify the condition \rlb{L>cond} for $L$ in \rlb{Lcheck} below.)
Recall that $\kPsi$ is an abbreviation for $\ket{\Psi_{(j_1,\ldots,j_N)}}$, the many-particle energy eigenstate labeled by $(j_1,\ldots,j_N)$.
Lemma~\ref{L:p<} guarantees that the bound \rlb{pLj<} is valid for at least $N-M$ distinct indices $j_i$.
For the rest, we use a trivial upper bound $p_q^{(j)}\le1$.
Since $\la\ge0$, we can upper bound the right-hand side of \rlb{PelaNP} to get
\eq
\bPsi e^{\la\hN_q}\kPsi\le
e^{M\la}\Bigl\{\Bigl(\frac{1}{m}+\delta\Bigr)\,e^\la+\Bigl(1-\frac{1}{m}-\delta\Bigr)\Bigr\}^{N-M},
\lb{PelaNP2}
\en
for any $N$ such that $N\ge M$.
(We shall check the condition $N\ge M$ in \rlb{Ncheck} below.)
The bound \rlb{PelaNP2} essentially recovers the rough estimate \rlb{PePrough} with an extra (but $L$-independent) factor $e^M$.

By substituting \rlb{PelaNP2} into \rlb{main3}, we find
\eqa
\bPsi\hP^+_q\kPsi&\le e^{M\la}\Bigl\{\Bigl(\frac{1}{m}+\delta\Bigr)\,e^\la+\Bigl(1-\frac{1}{m}-\delta\Bigr)\Bigr\}^{N-M}e^{-\la R}
\nl&\le\biggl(\frac{e^\la}{(\frac{1}{m}+\delta)\,e^\la+(1-\frac{1}{m}-\delta)}\biggr)^M
\Bigl\{\Bigl(\frac{1}{m}+\delta\Bigr)\,e^\la+\Bigl(1-\frac{1}{m}-\delta\Bigr)\Bigr\}^{N}e^{-\la R}
\nl&\le m^M
\Bigl\{\Bigl(\frac{1}{m}+\delta\Bigr)\,e^\la+\Bigl(1-\frac{1}{m}-\delta\Bigr)\Bigr\}^{N}e^{-\la R}
\nl&=m^M\Bigl\{f_{m,\delta}(\la)\,\exp\Bigl[-\la\Bigl(\frac{R}{N}-\frac{1}{m}\Bigr)\Bigr]\Bigr\}^N,
\lb{main4}
\ena
where we defined
\eq
f_{m,\delta}(\la)=
\Bigl(\frac{1}{m}+\delta\Bigr)\,e^{(1-\frac{1}{m})\la}+\Bigl(1-\frac{1}{m}-\delta\Bigr)e^{-\frac{1}{m}\la}.
\en
We can show that this function is bounded as
\eq
f_{m,\delta}(\la)\le e^{\la^2 A_2+\la\delta},
\lb{fbound}
\en
 for any $\la\in[0,1]$, with 
\eq
A_2=\frac{1}{m}\Bigl(1-\frac{1}{m}\Bigr)+\Bigl(1-\frac{2}{m}\Bigr)\delta.
\lb{C2}
\en
To see this, we expand $f_{m,\delta}(\la)$ in $\la$ as
\eq
f_{m,\delta}(\la)=\sum_{n=0}^\infty\frac{A_n}{n!}\la^n,
\en
with
\eq
A_n=\Bigl(\frac{1}{m}+\delta\Bigr)\Bigl(1-\frac{1}{m}\Bigr)^n+\Bigl(1-\frac{1}{m}-\delta\Bigr)\Bigl(-\frac{1}{m}\Bigr)^n.
\en
One finds that $A_0=1$, $A_1=\delta$, and $A_2$ is given by \rlb{C2}.
For $n\ge3$, we bound $A_n$ as
\eq
A_n\le\Bigl(\frac{1}{m}+\delta\Bigr)\Bigl(1-\frac{1}{m}\Bigr)^2+\Bigl(1-\frac{1}{m}-\delta\Bigr)\Bigl(\frac{1}{m}\Bigr)^2=A_2.
\en
Since $\la\in[0,1]$, we then have
\eq
f_{m,\delta}(\la)\le1+\la\delta+\la^2A_2\sum_{n=2}^\infty\frac{1}{n!}=1+\la\delta+\la^2A_2(e-2)
\le1+\la\delta+\la^2A_2,
\en
which implies \rlb{fbound}.

From \rlb{fbound}, we find
\eq
f_{m,\delta}(\la)\,\exp\Bigl[-\la\Bigl(\frac{R}{N}-\frac{1}{m}\Bigr)\Bigr]
\le\exp\Bigl[\la^2A_2-\la\Bigl(\frac{R}{N}-\frac{1}{m}-\delta\Bigr)\Bigr].
\lb{febound}
\en
We now set
\eq
\la=\frac{\frac{R}{N}-\frac{1}{m}-\delta}{2A_2},
\lb{lachoice}
\en
to minimize the right-hand side of \rlb{febound}.
Substituting the resulting bound to \rlb{main4}, we get
\eqa
\bPsi\hP^+_q\kPsi&
\le m^M\exp\biggl[-\frac{
(\frac{R}{N}-\frac{1}{m}-\delta)^2
}{4A_2}
 N\biggr]
\nl&
=m^M\exp\biggl[-\frac{
(\frac{R}{N}-\frac{1}{m}-\delta)^2
}{4\{\frac{1}{m}(1-\frac{1}{m})+(1-\frac{2}{m})\delta\}}
 N\biggr]
\nl&=m^M\exp\biggl[-\frac{
\{\frac{R}{L}-(\frac{1}{m}+\delta)\frac{N}{L}\}^2
}{4\{\frac{1}{m}(1-\frac{1}{m})+(1-\frac{2}{m})\delta\}}
 \frac{L}{N}L\biggr],
\lb{main5}
\ena
which essentially completes our estimate.
The remaining (tedious but straightforward) task is to simplify \rlb{main5}.

At this stage, we choose the small constant $\delta$ as
\eq
\delta=\frac{\eta}{m}.\lb{deltachoice}
\en
As we have noted already, this choice is quite arbitrary.
We can now verify the conditions for $L$ and $N$ mentioned above in the derivation of the key inequity \rlb{PelaNP2}.
Note that the assumption \rlb{L>} for $L$ now reads
\eq
L\ge\frac{3\epo}{u_0\delta}.
\lb{Lcheck}
\en
Since $u_0\le\epo+\eta$, the condition \rlb{L>cond} required in Lemma~\ref{L:p<} is safely satisfied.
We also note that \rlb{Lcheck} and \rlb{Nrange} imply
\eq
N\ge3\frac{1-\frac{\Duo}{u_0}}{1+\eta}\frac{1}{\delta}.
\lb{Ncheck}
\en
Noting that \rlb{u0small} and \rlb{Du0ep} imply
\eq
\eta\le\frac{1}{50},
\lb{etasmall}
\en
one can verify that  the right-hand side of \rlb{Ncheck} exceeds $\frac{2}{\delta}+\frac{1}{2}$.
This verifies the desired condition $N\ge M$.

By recalling the definition \rlb{Rdef} of $R$ and the upper bound for $N$ in \rlb{Nrange}, we have
\eqa
\frac{R}{L}-\Bigl(\frac{1}{m}+\delta\Bigr)\frac{N}{L}&\ge
\frac{u_0+\Du}{m(1+\eta)\epo}-\frac{1+\eta}{m}\,\frac{u_0+\Duo}{(1-\eta)\epo}
\nl&=\frac{1}{m(1+\eta)\epo}\Bigl\{
\Du-\frac{(1+\eta)^2}{1-\eta}\Duo-\frac{3\eta+\eta^2}{1-\eta}u_0\Bigr\}
\nl&\ge\frac{3}{4}\frac{1}{m(1+\eta)\epo}\Du.
\lb{main6}
\ena
We here noted that the assumed lower bound for $\Du$ in \rlb{Duconditions} and the upper bound \rlb{etasmall} for $\eta$ imply
\eq
\frac{3+\eta}{1-\eta}\eta u_0\le\frac{\Du}{5},\quad
\frac{(1+\eta)^2}{1-\eta}\Duo\le\frac{\Du}{20}.
\lb{bounds520}
\en
We also see that
\eqa
\frac{1}{m}\Bigl(1-\frac{1}{m}\Bigr)+\Bigl(1-\frac{2}{m}\Bigr)\delta
&=\frac{1}{m}\Bigl(1-\frac{1}{m}\Bigr)\biggl(
1+\frac{1-\frac{2}{m}}{\frac{1}{m}(1-\frac{1}{m})}\delta\biggr)
\nl&\le\frac{1}{m}\Bigl(1-\frac{1}{m}\Bigr)\{1+(m-2)\delta\}
\nl&\le(1+\eta)\frac{1}{m}\Bigl(1-\frac{1}{m}\Bigr),
\lb{mmbound}
\ena
where we recalled $m\delta=\eta$.

Substituting \rlb{main6} and \rlb{mmbound} and recalling the upper bound for $N$ in \rlb{Nrange}, the argument of the exponential function in \rlb{main5} is bounded as
\eqa
\frac{
\{\frac{R}{L}-(\frac{1}{m}+\delta)\frac{N}{L}\}^2
}{4\{\frac{1}{m}(1-\frac{1}{m})+(1-\frac{2}{m})\delta\}}
 \frac{L}{N}L
 &\ge
 \frac{1}{4\{(1+\eta)\frac{1}{m}(1-\frac{1}{m})\}}\Bigl(\frac{3}{4}\frac{1}{m(1+\eta)\epo}\Du\Bigr)^2
 \frac{(1-\eta)\epo}{u_0+\Duo}L
 \nl&=\frac{1}{4}\Bigl(\frac{3}{4}\Bigr)^2\frac{1-\eta}{(1+\eta)^3\bigl(1+\frac{\Duo}{u_0}\bigr)}\frac{(\Du)^2}{(m-1)\epo u_0}L
 \nl&\ge\frac{(\Du)^2}{8(m-1)\epo u_0}L.
 \lb{main7}
\ena
We here noted that the bounds \rlb{u0small} and \rlb{etasmall} for $\Duo/u_0$ and $\eta$ imply
\eq
\frac{1}{4}\Bigl(\frac{3}{4}\Bigr)^2\frac{1-\eta}{(1+\eta)^3\bigl(1+\frac{\Duo}{u_0}\bigr)}\ge\frac{1}{8}.
\lb{434bound}
\en
Substituting \rlb{main7} into \rlb{main5}, we finally get
\eq
\bPsi\hP^+_q\kPsi
\le m^{\frac{2m}{\eta}+\frac{1}{2}}\,\exp\Bigl[-\frac{(\Du)^2}{8(m-1)\epo u_0}L\Bigr],
\lb{main8}
\en
where we used the upper bound \rlb{Mcond} of $M$ and the choice \rlb{deltachoice} of $\delta$.
This is the desired \rlb{main01}.

We still need to justify the choice \rlb{lachoice} of $\la$ is in the range $[0,1]$.
That $\la\ge0$ was already shown in \rlb{main6}.
From \rlb{lachoice} and \rlb{Rdef}, we find
\eqa
\la&=\frac{\frac{R}{N}-\frac{1}{m}-\delta}{2\{\frac{1}{m}(1-\frac{1}{m})+(1-\frac{2}{m})\delta\}}
\le\frac{\frac{R}{N}-\frac{1}{m}}{2\,\frac{1}{m}(1-\frac{1}{m})}
\le\frac{mR}{N}-1
=\frac{u_0+\Du}{(1+\eta)\epo}\,\frac{L}{N}-1,
\ena
where we noted that $m\ge2$.
By using the lower bound for $N$ in \rlb{Nrange}, we see
\eq
\la\le\frac{u_0+\Du}{u_0-\Duo}-1=\frac{\Du-\Duo}{u_0-\Duo}\le1,
\en
where we noted $\Du\le u_0$ because of the assumed upper bound for $\Du$ in \rlb{Duconditions}.\footnote{%
We made a stronger assumption $\Du\le u_0/2$  in \rlb{Duconditions} since we need it in a similar evaluation for $\bPsi\hP^-_q\kPsi$.
}

The bound \rlb{main02} for $\bPsi\hP^-_q\kPsi$ is proved in a very similar manner.
Let us only sketch some parts that differ from the case for $\bPsi\hP^+_q\kPsi$.
The defining condition for $\hP^-_q$ is equivalent to $R'-\hN_q\ge0$ where $R'=\frac{u_0-\Du}{m(1-\eta)\epo}L$.
Corresponding to \rlb{main3} and \rlb{main4} above, we see, again for $\la\ge0$, that
\eqa
\bPsi\hP^-_q\kPsi&\le\bPsi e^{-\la\hN_q}\kPsi\,e^{\la R'}
\nl&\le\Bigl\{\Bigl(\frac{1}{m}-\delta\Bigr)\,e^{-\la}+\Bigl(1-\frac{1}{m}+\delta\Bigr)\Bigr\}^{N-M}e^{\la R'}
\nl&\le\biggl(\frac{1}{(\frac{1}{m}-\delta)\,e^{-\la}+(1-\frac{1}{m}+\delta)}\biggr)^M
\Bigl\{\Bigl(\frac{1}{m}-\delta\Bigr)\,e^{-\la}+\Bigl(1-\frac{1}{m}+\delta\Bigr)\Bigr\}^{N}e^{\la R'}
\nl&\le \Bigl(\frac{m}{m-1}\Bigr)^M
\Bigl\{\Bigl(\frac{1}{m}-\delta\Bigr)\,e^{-\la}+\Bigl(1-\frac{1}{m}+\delta\Bigr)\Bigr\}^{N}e^{\la R'}
\nl&=\Bigl(\frac{m}{m-1}\Bigr)^M\Bigl\{f_{m,-\delta}(-\la)\,\exp\Bigl[\la\Bigl(\frac{R'}{N}-\frac{1}{m}\Bigr)\Bigr]\Bigr\}^N,
\ena
We again have
\eq
f_{m,-\delta}(-\la)\le1+\la\delta+\la^2A'_2\le e^{\la\delta+\la^2A'_2},
\en
with $A_2'=\frac{1}{m}(1-\frac{1}{m})-(1-\frac{2}{m})\delta$.
It is crucial here to note that $\frac{R'}{N}-\frac{1}{m}\simeq-\Du/m\le0$.
Then we set
\eq
\la=-\frac{\frac{R'}{N}-\frac{1}{m}+\delta}{2A'_2}\ge0,
\lb{lachoice2}
\en
to minimize the right-hand side.
This leads to
\eq
\bPsi\hP^-_q\kPsi\le
\Bigl(\frac{m}{m-1}\Bigr)^M\exp\biggl[-\frac{
\{\frac{R'}{L}-(\frac{1}{m}-\delta)\frac{N}{L}\}^2
}{4\{\frac{1}{m}(1-\frac{1}{m})-(1-\frac{2}{m})\delta\}}
 \frac{L}{N}L\biggr],
\en
which corresponds to \rlb{main5}.
We get the desired \rlb{main02} after similar (but easier) estimates as above.

\paragraph*{Acknowledgement}
It is a pleasure to thank Shelly Goldstein, Hosho Katsura, Joel Lebowitz,  Shu Nakamura, Shin Nakano, Marcos Rigol, and Naoto Shiraishi for their useful discussions.
The present research is supported by JSPS Grants-in-Aid for Scientific Research No. 22K03474.



\begin{thebibliography}{10}

\bibitem{TasakiFreeFermion}
H. Tasaki,
{\em Macroscopic Irreversibility in  Quantum Systems: ETH and Equilibration in a Free Fermion Chain}\/,
(preprint, 2024).\\
\url{https://arxiv.org/abs/2401.15263}

\bibitem{vonNeumann}
J. von Neumann,
{\em Beweis des Ergodensatzes und des $H$-Theorems in der neuen Mechanik}\/,
Z. Phys. \textbf{57}, 30 (1929);\\
English translation (by R. Tumulka),
{\em Proof of the Ergodic Theorem and the H-Theorem in Quantum Mechanics}\/, The European Phys. J.  H {\bf 35} 201--237 (2010).\\
\url{https://arxiv.org/abs/1003.2133}

\bibitem{GLTZ}
S. Goldstein, J. L. Lebowitz, R. Tumulka, N. Zangh\`\i,
{\em Long-time behavior of macroscopic quantum systems: Commentary accompanying the English translation of John von Neumann's 1929 article on the quantum ergodic theorem}\/,
European Phys. J. H {\bf 35}, 173--200 (2010).\\
\url{https://arxiv.org/abs/1003.2129}

\bibitem{GLMTZ09b} 
S. Goldstein, J. L. Lebowitz, C. Mastrodonato, R. Tumulka, and N. Zangh\`\i,
{\em On the Approach to Thermal Equilibrium of Macroscopic Quantum Systems}\/,
Phys. Rev. E \textbf{81}, 011109 (2010).\\
\url{https://arxiv.org/abs/0911.1724}

\bibitem{Tasaki2010}
H. Tasaki,
{\em The approach to thermal equilibrium and ``thermodynamic normality" --- An observation based on the works by Goldstein, Lebowitz, Mastrodonato, Tumulka, and Zanghi in 2009, and by von Neumann in 1929}\/, (unpublished note 2010).
\\\url{https://arxiv.org/abs/1003.5424}

\bibitem{Tasaki2016}
H. Tasaki,
{\em Typicality of thermal equilibrium and thermalization in isolated macroscopic quantum systems}\/,
J. Stat. Phys. {\bf 163}, 937--997 (2016).
\\\url{https://arxiv.org/abs/1507.06479}



\bibitem{Deutsch1991}
J.M. Deutsch,
{\em Quantum statistical mechanics in a closed system}\/,
Phys. Rev. A {\bf 43}, 2046 (1991).

\bibitem{Srednicki1994}
M. Srednicki,
{\em Chaos and quantum thermalization}\/,
Phys. Rev. E {\bf 50}, 888 (1994).

\bibitem{Tasaki1998}
H. Tasaki,
{\em 	From Quantum Dynamics to the Canonical Distribution: General Picture and a Rigorous Example}\/, 
Phys. Rev. Lett.  \textbf{80}, 1373--1376 (1998).\\
\url{https://arxiv.org/abs/cond-mat/9707253}

\bibitem{RigolSrednicki2012}
M. Rigol and M. Srednicki,
{\em Alternatives to Eigenstate Thermalization}\/,
Phys. Rev. Lett. {\bf 108}, 110601 (2012).
\\\url{https://arxiv.org/abs/1108.0928}

\bibitem{DAlessioKafriPolkovnikovRigol2016}
L. D'Alessio, Y. Kafri, A. Polkovnikov, and M. Rigol,
{\em From quantum chaos and eigenstate thermalization to statistical mechanics and thermodynamics}\/,
Adv. Phys. {\bf 65},  239--362 (2016).
\\\url{https://arxiv.org/abs/1509.06411}




\bibitem{GoldsteinHuseLebowitzTumulka2015}
S. Goldstein, D.A. Huse, J.L. Lebowitz, and R. Tumulka,
{\em Thermal equilibrium of a macroscopic quantum system in a pure state}\/,
Phys. Rev. Lett. {\bf 115}, 100402 (2015).
\\\url{https://arxiv.org/abs/1506.07494}




\bibitem{ShiraishiTasaki2023}
N. Shiraishi and H. Tasaki,
{\em Nature abhors a vacuum: A simple rigorous example of thermalization in an isolated macroscopic quantum system}\/,
(preprint, 2023).
\\\url{https://arxiv.org/abs/2310.18880}



\bibitem{GluzaEisertFarrelly2019}
M. Gluza, J. Eisert, and T. Farrelly,
{\em Equilibration towards generalized Gibbs ensembles in non-interacting theories}\/,
SciPost Phys. {\bf 7}, 038 (2019).
\\\url{https://www.scipost.org/10.21468/SciPostPhys.7.3.038}

\bibitem{RigolMuramatsuOlshanii2006}
M. Rigol, A. Muramatsu, and M. Olshanii,
{\em Hard-core bosons on optical superlattices: Dynamics and relaxation in the superfluid and insulating regimes}\/,
Phys. Rev. A {\bf 74}, 053616 (2006).
\\\url{https://arxiv.org/abs/cond-mat/0612415}

\bibitem{RigolFitzpatrick2011}
M. Rigol and M. Fitzpatrick,
{\em Initial state dependence of the quench dynamics in integrable quantum systems}\/,
Phys. Rev. A {\bf 84}, 033640 (2011).
\\\url{https://arxiv.org/abs/1107.5811}

\bibitem{Pandeyetal}
S. Pandey, J.M. Bhat, A. Dhar, S. Goldstein, D.A. Huse, M. Kulkarni, A. Kundu, and J.L. Lebowitz,
{\em Boltzmann entropy of a freely expanding quantum ideal gas}\/,
J. Stat. Phys. {\bf 190}, article number 142, (2023).
\\\url{https://arxiv.org/abs/2303.12330}







\bibitem{Hal_NT}
H. Tasaki, {\em Two number-theoretic theorems (that we found useful for quantum physics) and their elementary proofs}\/, YouTube video 2023.
\\\url{https://youtu.be/YrCoBv0acgs}

\bibitem{Tignol}
J.-P. Tignol,
{\em Galois' Theory of Algebraic Equations (second edition)}\/, (World Scientific, 2015).

\bibitem{IR}
K. Ireland and M. Rosen,
{\em A Classical Introduction to Modern Number Theory}\/, (Graduate Texts in Mathematics, Springer, 1990).


\bibitem{StewarTall}
I. Stewart and D. Tall,
{\em Algebraic number theory and Fermat's last theorem}\/, (Chapman and Hall, 2020).







\end{thebibliography}
\end{document}